\documentclass[pra,twocolumn,showpacs,amsmath,amssymb,nofootinbib,floatfix]{revtex4}
\input epsf
\begin{document}
%\draft
\date{\today}

\title{Bohmian Histories and Decoherent Histories}

\author{James B.  Hartle \thanks{hartle@physics.ucsb.edu}}

\vskip .13 in
\affiliation{Department of Physics, University of California,
Santa Barbara, CA
93106-9530}

\begin{abstract}
%\tighten

The predictions of the Bohmian and the decoherent (or consistent) histories
formulations of the quantum mechanics of a closed system are compared for
{\it histories} --- sequences of alternatives at a series of times.
For certain kinds of histories, Bohmian
mechanics and decoherent histories may both be formulated in the same
mathematical framework within which they can be compared. 
In that framework,
Bohmian mechanics and decoherent histories represent a given
history by different operators. Their predictions for the probabilities of
histories of a closed system therefore generally differ. However, in an 
idealized model of
measurement, the predictions of Bohmian mechanics and decoherent histories
coincide for the probabilities of records of
measurement outcomes. The formulations are thus difficult to distinguish
experimentally. They may differ in their accounts of the past
history of the universe in quantum cosmology. 
\end{abstract}


\maketitle
%\tighten
%\narrowtext
\setcounter{footnote}{0}
\section{Introduction}
\label{sec: I}

Bohmian mechanics ({\it e.g.} \cite{BH93}) and decoherent (or consistent)
histories quantum mechanics ({\it e.g.} \cite{Gri84,Omn94,GH90a}) 
are formulations of
quantum mechanics that share some common objectives. Both are
formulations of quantum mechanics for a closed system such as the universe.
Both are formulations of quantum mechanics that do not posit a
privileged, fundamental role for measurements or the observers that make
them. The question of the relationship between these two formulations
has been addressed in several different
places \cite{BH93,Ken96,Gri99,HM00}. This paper compares the predictions of the
two formulations for {\it histories} --- time sequences of alternatives for the 
closed system.
We make the following points:

\begin{itemize}

\item For certain classes of histories, Bohmian mechanics and decoherent histories (for short) may both be
formulated within the same mathematical framework of a Hilbert space of
states, operators representing alternatives, and unitary evolution. Within
this framework they may be compared. 

\item Bohmian mechanics and decoherent histories generally represent the
same history by different operators.  Their predictions for the
probabilities of histories will therefore generally differ.

\item In an idealized model of measurement, Bohmian
mechanics and decoherent histories predict the same probabilities
for records of the
outcomes of measurements. This makes
it difficult to distinguish the formulations experimentally.
\end{itemize}

Not surprisingly, there is overlap of our comparison of Bohmian
mechanics and decoherent histories with others 
especially that of Griffiths {\cite{Gri99}. However,
we do not aim in this paper to present arguments for or against either
formulation of the quantum mechanics of closed, non-relativistic systems.
We intend to merely point out some differences between them.

After a necessary review of the two formulations in Section II, we
establish the above three facts in Sections III--V. We conclude with some 
brief discussion of possible differences between the descriptions
given by Bohmian mechanics and decoherent histories of the past in
cosmology.  
\section{Bohmian Mechanics and Decoherent Histories}
\label{sec: II}

\subsection{A Model Closed System}

To establish the facts listed in the Introduction, it is sufficient to use
the illustrative (but unrealistic) model of a universe of $N$
non-relativistic particles in a box. The dynamics of the positions $\vec
x_1, \cdots, \vec x_N$ is governed by a Hamiltonian assumed to be of the
form 
\begin{equation}
H = \sum^N_{i=1}\ \frac{\vec p^{\, 2}_i}{2m_i} + V(\vec x_1, \cdots, \vec
x_n)\ .
\label{twoone}
\end{equation}
In addition to the Hamiltonian $H$, we assume that the initial quantum state of
the closed system $|\Psi\rangle$ is given at time $t=t_0=0$. Its configuration 
space
representative is the initial wave function
\begin{equation}
\Psi \left(\vec x_i, 0\right) \equiv \left\langle \vec x_1, \cdots \vec x_n
|\Psi\right\rangle\ .
\label{twotwo}
\end{equation}

The most general objective of a quantum mechanical theory of  a closed
system is the prediction of the probabilities of the individual members of a
set of alternative, coarse-grained, time histories of the system. For
example, if the box contained the solar system, probabilities of 
different orbits of the earth around the sun might be of interest. These
orbits
are histories of the position of the center of mass of the earth at a
sequence of times. A coarse-grained history might be specified by giving a
sequence of ranges $\Delta_1, \cdots, \Delta_n$ for the center of mass
position of the earth at a series of times $t_1, \cdots, t_n$. The history
is coarse grained because the position of every particle in the box 
is not specified,
the center of mass position is not specified to arbitrary accuracy, and not
at all possible times.  Bohmian mechanics is usually formulated in terms of
histories of position, so it will be convenient to restrict the further
discussion to histories of this type. Specifically, we consider histories
specified by giving exhaustive sets of exclusive regions
$\{\Delta^k_{\alpha_k}\}$, $\alpha_k=1, 2, \cdots$, of the configuration
space of the $\vec x_i$'s at a series of times $t_k$, $k=1, \cdots, n$. A
history is thus specified by a series of regions $(\alpha_n, \cdots,
\alpha_1)$ which we denote by $\alpha$ for short.

\subsection{Decoherent Histories Quantum Mechanics}

We now briefly review how (and when) decoherent histories quantum mechanics
assigns probabilities to a history $\alpha\equiv (\alpha_n, \cdots,
\alpha_1)$ of ranges of position at a series of times $t_1, \cdots, t_n$.
For more details in the present notation see, 
{\it e.g.} \cite{GH90a,Har93a}. 

The alternative regions of configuration space
$\{\Delta^k_{\alpha_k}\}$ correspond to
an exhaustive set of exclusive (Schr\"odinger picture) projection operators $\{P^k_{\alpha_k}\}$
that project onto these regions. Because the regions are exhaustive and
exclusive these projection operators satisfy (for each $k$)
\begin{equation}
P^k_{\alpha_k} P^k_{\alpha^\prime_k} = \delta_{\alpha_k\alpha^\prime_k}
P^k_{\alpha_k} \quad , \quad \sum\nolimits_{\alpha_k} P^k_{\alpha_k}
= I\ .
\label{twothree}
\end{equation}
A history of alternatives $\alpha=(\alpha_n, \cdots, \alpha_1)$ at times
$t_1, \cdots, t_n$ is represented by the corresponding chain of projections
interspersed with unitary evolution
\begin{equation}
C_\alpha \equiv P^n_{\alpha_n} U \left(t_n, t_{n-1}\right)
P^{n-1}_{\alpha_{n-1}} U \left(t_{n-1}, t_{n-2}\right) \cdots 
P^1_{\alpha_1} U\left(t_1, 0\right)\ ,
\label{twofour}
\end{equation}
where
\begin{equation}
U\left(t^{\prime\prime}, t^\prime\right) =
e^{-iH\left(t^{\prime\prime}-t^\prime\right)/\hbar}\ ,
\label{twofive}
\end{equation}
and $t=t_0=0$ is the time of the initial condition.

The probabilities $p_\alpha$ of the individual histories in a set of
alternative histories are given by
\begin{equation}
p^{(DH)}_\alpha = \left\Vert C_\alpha |\Psi \rangle \right\Vert^2\ .
\label{twosix}
\end{equation}
However, decoherent histories quantum mechanics does not assign
probabilities to every set of histories that may be described. The numbers
(\ref{twosix}) may be inconsistent with the rule that the probability of a
coarser-grained set of alternatives should be the {\it sum} of
probabilities of its members. Rather, probabilities are assigned only to
sets of alternative histories that  are consistent
\cite{Gri84}, for example by satisfying the decoherence condition
\begin{equation}
\langle \Psi | C^\dagger_{\alpha^\prime} C_\alpha |\Psi\rangle
\approx 0\quad {\rm for}\ \alpha^\prime \not= \alpha\ .
\label{twoseven}
\end{equation}

We stress that ``decoherence'', ``decoherent'' , etc. as used in
this paper refer to the absence of interference between the {\it
histories} in an exhaustive set of alternative histories as specified
quantitatively by (\ref{twoseven}). We do not mean a process in
which a reduced density matrix becomes approximately diagonal  
which is another common usage of these terms\footnote{Although 
not particularly relevant for this paper, a discussion of the
connections and differences between these two usages for
``decoherence'' can be found in the ``note added'' to \cite{GH94}.}. 

\subsection{Bohmian Mechanics}

In Bohmian mechanics, the trajectories of the particles in the box obey two
deterministic equations. The first is the Schr\"odinger equation for
$\Psi$.
\begin{subequations}
\begin{equation}
i\hbar\ \frac{\partial\Psi}{\partial t} = H\Psi\ .
\label{twoeight a}
\end{equation}
Then, writing $\Psi = R\exp(iS)$ with $R$ and $S$ real, the second is the
deterministic equation for the $\vec x_i(t)$
\begin{equation}
m_i\ \frac{d\vec x_i}{dt} = \vec \nabla_{\vec x_i} S\left(\vec x_1,
\cdots, \vec x_N\right)\ ,
\label{twoeight b}
\end{equation}
\label{twoeight}
\end{subequations}
The initial wave function (\ref{twotwo}) is the initial condition for 
(\ref{twoeight
a}). The theory becomes a statistical theory with the
assumption that the {\it initial values} of the $\vec x_i$ are distributed
according to the probability density on configuration space
\begin{equation}
\wp\left(\vec x_1, \cdots, \vec x_N\, , 0\right) = \left|\Psi\left(\vec x_1,
\cdots, \vec x_N, 0\right)\right|^2\ .
\label{twonine}
\end{equation}
Once this initial probability distribution is fixed, the probability of any
later alternatives is fixed by the deterministic equations
(\ref{twoeight}).

A coarse-grained Bohmian history $\alpha \equiv (\alpha_n, \cdots,
\alpha_1)$ defined by a sequence of ranges $\{\Delta^k_{\alpha_k}\}$
 of the $\vec x_i$ at a series of
times consists of the set of Bohmian trajectories $\vec
x_i(t)$ that cross those ranges at the specified times.

\section{Bohmian and Decoherent Histories}
\label{sec: III}

\subsection{Bohmian Histories}

An individual
Bohmian trajectory $\vec x_i(t)$ is fixed deterministically by equations
(\ref{twoeight}) once the initial condition $\vec x_i(0)$ and the initial
value of $\Psi$ are given. The
probability of a coarse-grained history $\alpha \equiv (\alpha_n, \cdots,
\alpha_1)$ is therefore the probability of the region of initial $\vec
x_i(0)$'s that lead to trajectories that pass through the regions
$\Delta^n_{\alpha_n}, \cdots, \Delta^1_{\alpha_1}$ at times $t_1, \cdots,
t_n$. We call this range  of initial values $\Delta^{(BM)}_{\alpha_1 \cdots \alpha_n}$ or
$\Delta^{(BM)}_\alpha$ for short.
Denote by $B_\alpha$ the projection on this range of $\vec x_i(0)$'s
corresponding to the history $\alpha$.  From (\ref{twonine}) the probability
$p^{(BM)}_\alpha$ predicted by Bohmian mechanics for this history is
\begin{equation}
p^{(BM)}_\alpha = \Vert B_\alpha | \Psi \rangle \Vert^2
\ .
\label{threeone}
\end{equation}
The operator $B_\alpha$ is not determined by the ranges
$\Delta^n_{\alpha_n}, \cdots, \Delta^1_{\alpha_1}$ alone but also
depends\footnote{The author owes this observation to T. Erler.}  on the initial state $\Psi$. That is because the evolution
of $\Psi$ through (\ref{twoeight b}) is needed to determine  whether
the trajectories pass through the regions $\Delta^n_{\alpha_n},
\cdots, \Delta^1_{\alpha_1}$. 

\subsection{Different Probabilities for the Same History}

The Bohmian formula (\ref{threeone}) is an expression for the probabilities 
of a
set of histories that is in the same mathematical framework as
(\ref{twosix}) for decoherent histories. It is thus possible to compare
the two formulations.

\begin{figure}[t]
\centerline{\epsfysize=3.00in \epsfbox{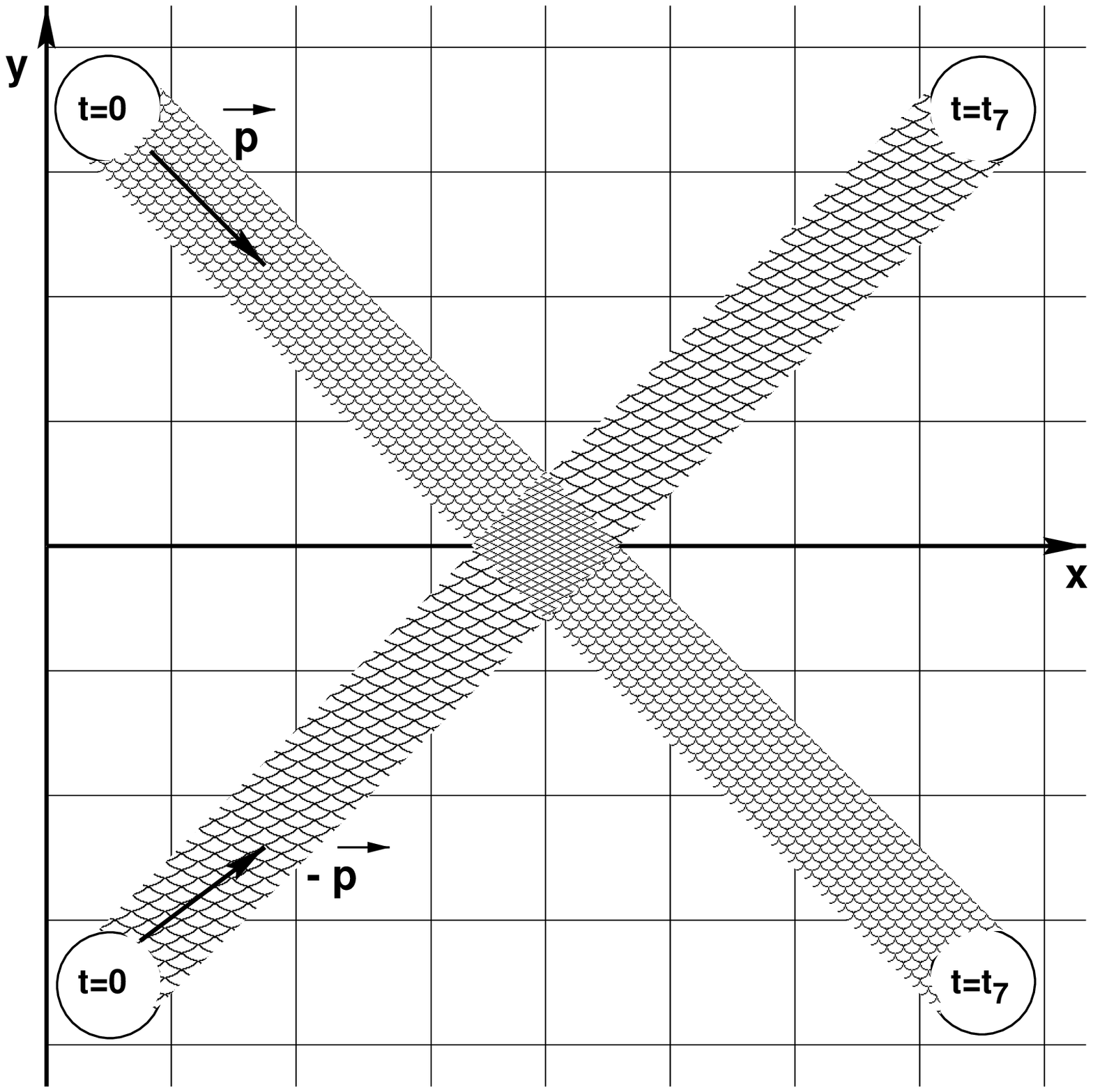}}
\caption{\sl
The motion of wave packets in the BESSW example. Two wave packets
initially localized (at $t=0$) within the circles at left  have equal and
opposite $y$-components of their expected momentum $\vec p$. The wave
packets evolve through the shaded regions indicated to the positions
shown at $t=t_7$. At intermediate times $t_1, \cdots t_6$, they are
centered in
one or the other of the squares.}
\end{figure}

We first note that decoherence in the sense of the absence of 
interference between multi-time histories [{\it cf} (\ref{twoseven})]
is automatic for sets of Bohmian histories.
The $B_\alpha$ are orthogonal
projections onto disjoint regions of the initial $\vec x_i(0)$ and satisfy
$B_\alpha B_{\alpha^\prime} = \delta_{\alpha \alpha^\prime}B_\alpha$ [{\it
cf.} (\ref{twothree})]. Thus, 
\begin{equation}
\langle \Psi \left| B^\dagger_\alpha B_{\alpha^\prime} \right| \Psi
\rangle \equiv 0 \quad , \quad \alpha\not= \alpha^\prime\ .
\label{threetwo}
\end{equation}
Consistency of Bohmian probabilities is automatic.

This is the first of the differences between the Bohmian mechanics and
the decoherent histories formulations of quantum theory: {\it Bohmian
mechanics assigns probabilities to sets of histories of position that 
do not decohere.} Bohmian mechanics therefore potentially assigns 
probabilities to more sets of histories of position than decoherent
histories. However, when histories of alternatives other than
position are considered, the situation is the other way around. 
The decoherent histories formulation assigns probabilities to histories of
alternatives defined by ranges of operators other than position
which are not represented, at least not fundamentally, in Bohmian
mechanics. 

A comparison of the probability expressions (\ref{threeone}) and
(\ref{twosix}) reveals a second and more important difference. In general, {\it Bohmian mechanics and
decoherent histories predict different probabilities for the same
history}. That is because the operator $B_\alpha$ corresponding to a
history in Bohmian mechanics is a projection while the operator $C_\alpha$
is decoherent histories is generally not.\footnote{This is the case even
though the $C_\alpha$ have certain similarities to projections for {\it
decoherent sets, e.g.} 
\[
p^{(DH)}_\alpha = \langle\Psi | C^\dagger_\alpha C_\alpha | \Psi\rangle =
\left\langle\Psi\left|C_\alpha\right|\Psi\right\rangle
\]
}
We shall give explicit examples below.

We should
perhaps stress that we are not referring here to histories of
measurements of position carried out by some external
system. We are rather considering both Bohmian mechanics and
decoherent histories as quantum theories of a closed system
containing measurement apparatus if any. The special situation
with histories of measured alternatives will be discussed in Section
IV.

The one general case when the probabilities coincide are when the set of
histories is so coarse grained that it consists of alternatives 
$\{P_\alpha\}$ at a single moment of time $t_1$
\begin{equation}
C_\alpha =  P_\alpha U\left(t_1, 0\right)\ .
\label{threethree}
\end{equation}
Probability is conserved along Bohmian trajectories. Specifically it
follows from (\ref{twoeight}) that
\begin{equation}
\frac{\partial \wp}{\partial t} + \sum^N_{i=1} \vec\nabla_{\vec x_i} \cdot
\left(\wp\ \frac{\nabla_{\vec x_i} S}{m_i}\right) = 0
\label{threefour}
\end{equation}
where $\wp=\left| \Psi \left(\vec x_i, t\right) \right|^2$. Therefore, since
$B_\alpha$ are projections on the range of positions at $t=0$ of
trajectories that pass
through the regions defined by $P_\alpha$ at time $t$, 
\begin{equation}
\left\Vert P_\alpha U \left(t_1, 0\right) | \Psi \rangle  \right\Vert^2 =
\left\Vert B_\alpha |\Psi\rangle \right\Vert^2\ ,
\label{threefive}
\end{equation}
and 
\begin{equation}
p_\alpha^{(DH)} = p^{(BM)}_\alpha \quad , \quad ({\rm for\ single\ time
\ histories)}\ .
\label{threesix}
\end{equation}

The time $t_1$ is arbitrary. Probabilities of histories restricted to a
single time coincide for all values of that time. But most physically
interesting histories are described by alternatives at more than one
time. For example, predictions of the orbit of the Mars around the 
Sun involve the conditional probabilities for future positions of the
Mars given some observations of its position in the past. Those 
are constructed from the probabilities of histories of the location
of Mars at multiple times in the past and future. Indeed, in the
context of quantum cosmology,  very few useful predictions can be 
expected from alternatives at a single time conditioned only by
the initial state of the universe $|\Psi\rangle$ (see, {\it e.g.}
\cite{Har03a}).  

\subsection{An Example}

Several situations that have been widely discussed in Bohmian mechanics
provide examples where the probabilities predicted by Bohmian mechanics
differ significantly from those of decoherent histories.
\cite{Bel80,Engsum,DFGZ93,DHS93,AV96,HCM00,Gri99}. We call these
BESSW examples.
A simple case is illustrated in Figure 1.

A single free particle moves in a two-dimensional plane. The plane is
divided into square regions
 that will be used to describe coarse-grained histories
of the particle's position.  (The squares are the $\{\Delta^k_{\alpha_k}\}$ 
with
the same set of ranges for each $k$.) The initial wave function is a
superposition of two wave packets $G(x,y, 0)$ and $G(x,-y,0)$, {\it viz.},
\begin{equation}
\Psi (x,y, 0) = \frac{1}{\sqrt{2}}\ \left[G(x,y, 0) + G(x, -y, 0)\right]\ .
\label{threeseven}
\end{equation}
The wave packet $G(x,y, 0)$ is assumed to be localized well within the
dimensions of one initial square as shown, and to have a momentum $\vec p$
defined within the limits of the uncertainty principle with a negative 
component
of $p_y$. The wave packet $G(x,-y, 0)$ is initially located symmetrically about
the $y$ axis and has an equal magnitude but positive  component of $p_y$. Evolved by the
Schr\"odinger equation over a time interval short compared to that for
significant spreading,  the wave packets will move through the shaded
regions shown in Figure 1. 

To define a set of alternative histories, choose a sequence of equally
spaced times $t_1, t_2, \cdots, t_7$ when each wave packet is within
one of the squares encompassing its path. A set of all possible
coarse-grained histories
is then the set of all possible sequences of squares at these times. We now
calculate the probabilities assigned by Bohmian mechanics and decoherent
histories to this set of alternative coarse-grained histories.
Our analysis does not differ substantially from that presented
by Griffiths \cite{Gri99} for the same class of examples. 

The probabilities predicted by decoherent histories are given by
(\ref{twosix}) and (\ref{twofour}). The action of the operator $C_\alpha$
can be described as successive projections (or ``reductions'') on the
sequence of squares defining the history interspersed with unitary
evolution. 
The unitary evolution moves the wave packet as described above.
Since the wave packets are within one square or another at the times $t_1,
\cdots, t_7$, the projections have almost no effect on them. Thus, for the
history $\alpha=\alpha_+$ corresponding to the sequence of squares that track the wave
packet starting at upper left
\begin{subequations}
\begin{equation}
\langle x,y  | C_{\alpha_+}|\Psi\rangle \approx 
\frac{1}{\sqrt{2}}\ G\left(x,y,t_7\right)\ .
\label{threeeight a}
\end{equation}
Similarly for the $\alpha=\alpha_-$ history corresponding to the squares that track
the wave packet starting at lower left
\begin{equation}
\langle x,y |C_{\alpha_-}|\Psi\rangle \approx
\frac{1}{\sqrt{2}}\ G\left(x, -y, t_7\right)\ .
\label{threeeight b}
\end{equation}
For all other sequences of squares
\begin{equation}
\langle x,y\left|C_\alpha\right| \Psi\rangle\approx 0 \ , \quad \alpha
\not= \alpha_\pm\ .
\label{threeeight c}
\end{equation}
\label{threeeight}
\end{subequations}
These facts imply that the set of histories is approximately decoherent
because the only potentially non-zero, off-diagonal inner product is
\begin{equation}
\langle\Psi\left|C^\dagger_{\alpha_+}
C_{\alpha_-}\right|\Psi\rangle\approx 0 \ .
\label{threenine}
\end{equation}
This  is negligible because there is almost no overlap between the two wave
packets at $t_7$. The only non-vanishing probabilities are for the
sequences $\alpha_+$ and $\alpha_-$ with
\begin{equation}
p^{(DH)}_{\alpha_+} = p^{(DH)}_{\alpha_-} \approx \frac{1}{2}\ .
\label{threeten}
\end{equation}
These two histories are illustrated in Figure 2.
\begin{figure}[t]
\centerline{\epsfysize=3.00in \epsfbox{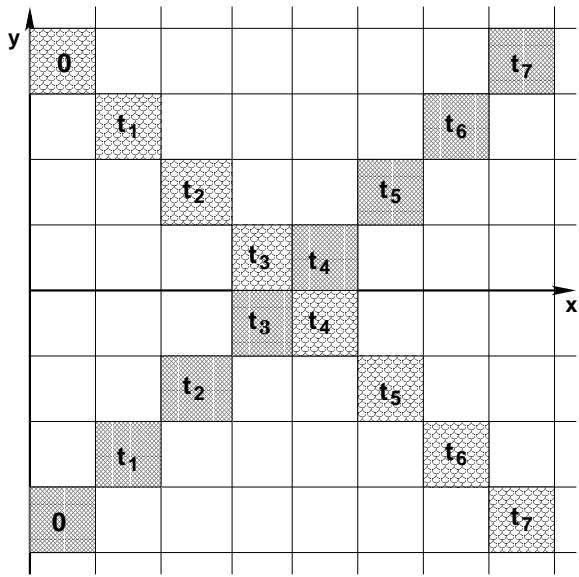}}
\caption
{\sl Coarse-grained histories with high probability in the BESSW
example according to decoherent histories quantum mechanics. Coarse-grained
histories are defined by the grid of square regions at a sequence of times
$t_1, \cdots, t_7$ adjusted so that the wave packets of Fig.~1 are centered
in one square at each time. The possible coarse-grained histories are the
sequence of any seven squares at the series of times. The two sequences of
squares which have any significant probability according to decoherent
histories quantum mechanics are shown, distinguished by different
shadings.}
\end{figure}

We now turn to the predictions of Bohmian mechanics for the same set of
histories. Because $\Psi(x,y, 0)$ is symmetric about the
$x$-axis, and because the Hamiltonian commutes with this symmetry, it holds
for all times when $\Psi(x,y,0)$ is evolved by the Schr\"odinger equation
(\ref{twoeight a})
\begin{equation}
\Psi(x,y,t)=\Psi(x, -y, t)\ .
\label{threeeleven}
\end{equation}
Then from (\ref{twoeight b})
\begin{equation}
\left(\frac{dy}{dt}\right)_{x=0} = 0
\label{threetwelve}
\end{equation}
for all times. No Bohmian trajectory crosses the $x$-axis. In particular, 
\begin{figure}[t]
\centerline{\epsfysize=3.00in \epsfbox{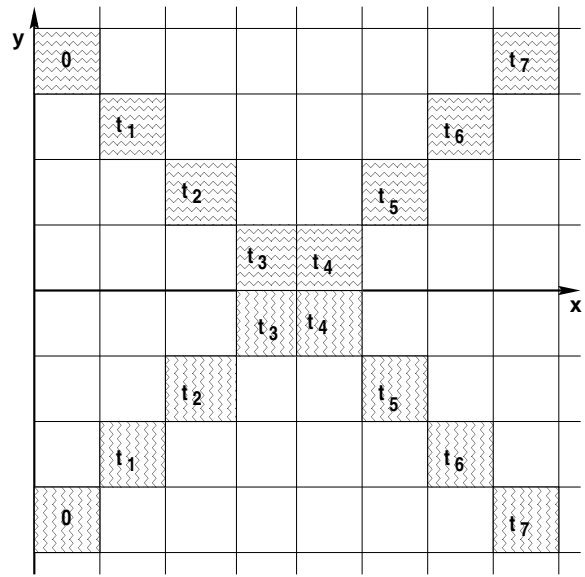}}
\caption
{\sl Coarse-grained histories with high probability in the BESSW
example according to Bohmian mechanics. The coarse-grained histories are
the same as those in Fig.~2 ({\it q.v.}). The two sequences of squares with
any
significant probability are shown, distinguished by different
shadings. They are different from those predicted
by decoherent histories. }
\end{figure}

\begin{equation}
p^{(BM)}_{\alpha_+} = p^{(BM)}_{\alpha_-} \approx  0\ .
\label{threethirteen}
\end{equation}
The two histories for which Bohmian mechanics predicts non-negligible
probabilities are sketched in Figure 3. However, only a comparison of
(\ref{threeten}) and (\ref{threethirteen}) is necessary to demonstrate
conclusively that decoherent histories and Bohmian mechanics generally
predict different probabilities for the same set of alternative
coarse-grained histories of a closed system. 

The reason for the difference is also illustrated by this example. We
mentioned that the action of the $C_\alpha$ could be thought of as unitary
evolution interspersed with reduction. But in Bohmian mechanics, the wave
function is never reduced. It evolves on forever by the Schr\"odinger
equation undisturbed by any ``second law of evolution''.

An important point illustrated by the BESSW example is that Bohmian
trajectories while always deterministic are not necessarily {\it
classically} deterministic. Quantum effects can be important for Bohmian 
trajectories even
in situations where (as here) wave packets move approximately classically. 

\subsection{Another Viewpoint}

Mathematically, the different predictions of Bohmian mechanics and
decoherent histories arise because the same history $\alpha$ is
represented by different operators --- $B_\alpha$ and $C_\alpha$
respectively. However, from the decoherent histories viewpoint,
the $B_\alpha$ describe a set of {\it single time}
histories of various ranges of position $\Delta^{(BM)}_\alpha$
at $t=0$. (Recall the definition of $B_\alpha$ in Section \ref{sec:
III}.A.) 
That is, the operator $B_\alpha$ describes a history in the decoherent
histories formulation of quantum mechanics ---  not a sequence of alternatives
at a {\it series} of times, but a certain range of positions at {\it one} time.
Mathematically, therefore Bohmian mechanics is a restriction of decoherent
histories quantum mechanics to histories represented by operators of a
particularly simple type. Within decoherent histories quantum
mechanics, 
the range $\Delta^{(BM)}_\alpha$ could be described as ``the range of
initial positions which, if evolved by the equations of Bohmian mechanics, 
would lead
to trajectories passing through the $\{\Delta^k_{\alpha_k}\}$ at times
$t_k$''. Bohmian mechanics would thus be employed merely as a tool for
describing certain single time histories in decoherent histories quantum
mechanics. However, to adopt this position would be tendentious. We will
take the alternative view that Bohmian mechanics is not merely a different way of
describing operators, but is a different way of interpreting them. In
particular we will take the view that Bohmian mechanics and decoherent
histories represent the same history by different operators and therefore
may be different predictions for their probabilities.
In the next section we discuss whether these can be
observed.

\section{Measurements}
\label{sec: IV}

Are the different predictions of Bohmian mechanics and decoherent
histories testable by experiments? That depends on whether the two
formulations predict different probabilities for the outcomes of
measurements. In this section we analyze this question employing
idealizations common in many measurement models. 
(See, {\it e.g.} \cite{Har91a}.)

Both Bohmian mechanics and decoherent histories are formulations of
quantum mechanics for a closed system that do not posit a fundamental role
for measurements or observers. Measurements and observation can, of
course, be described in a closed system that contains both observer and
observed, both measured subsystem and measurement apparatus. With suitable
idealizations, the usual results of the approximate quantum mechanics of
measured subsystems (Copenhagen quantum mechanics) are recovered to an
excellent approximation \cite{Har91a}.

Various general characterizations of measurement have been proposed --- an
``irreducible act of amplification'', ``correlation of a
`microscopic' variable
with a `macroscopic' variable'', {\it etc.}  It has proved difficult to make such
ideas precise ({\it e.g.} what is ``macroscopic''?), but a precise and
general characterization is not needed in a quantum mechanics of a closed
system.  One characteristic which seems generally agreed upon 
is that the results of a measurement must be recorded --- at least for 
a time.  Histories of measurements whose outcomes are recorded can be 
modeled as follows in a closed system containing both measurement 
apparatus and measured subsystem.

For simplicity consider a single apparatus which carries out a
series of measurements on another subsystem over a series of times.
The apparatus records the sequence of outcomes for examination at
some time $t_R$ after all the measurements are completed. Let
$\{R_\alpha\}$ be set of orthogonal projection operators describing
the alternative values of these records at $t_R$. To preserve
the contact with Bohmian mechanics we shall assume that the $R$'s
are projections onto ranges of the $x's$, as is plausibly the case
for records of many realistic measurement situations. 

Bohmian mechanics and decoherent histories will agree on the
predictions of probabilities for the alternative outcomes of the 
measurements registered in these records. That is because the
$\{R_\alpha\}$ represent alternatives at a single time whose
probabilities generally agree as discussed in the last section
[{\it cf} (\ref{threesix})]. Thus, if the result of every experiment can be
summarized in records that are coarse-grained alternatives of the 
positions $\vec x_i$ at a single time, there seems little prospect that 
experiment can distinguish Bohmian mechanics from decoherent histories.   

However, even if there is agreement on the probabilities of the
records there can be disagreement on what they record, or indeed
whether they are records at all, in situations where Bohmian
mechanics and decoherent histories disagree on the probabilities of
histories. To understand this let us first review more precisely what it
means for a set of projection operators $\{R_\alpha\}$ to be a record
of a history beginning with the case of decoherent histories.

Let $C_{\alpha_n \cdots \alpha_1}$ be the operator representing a
history of coarse-grained alternatives that have been measured, and let
$R^{(C)}_{\beta_n \cdots \beta_1}$ denote the orthogonal projections
onto the various possible values of a record of the measurements.
(The superscript ``$C$'' stands for ``records of the $C$'s'' --- not
``consistent''.)  In an
ideal measurement situation, the records at a later time $t_R$ are exactly
correlated with the history of measured alternatives, {\it viz.}
\begin{equation}
R^{(C)}_{\beta_n \cdots \beta_1} e^{-iH(t_R-t_n)/\hbar}
C_{\alpha_n\cdots \alpha_1} |\Psi\rangle \propto \delta_{\beta_n\alpha_n}
\cdots \delta_{\beta_1\alpha_1}\ ,
\label{fourone}
\end{equation}
a relation we abbreviate by
\begin{equation}
R^{(C)}_\beta e^{-iH(t_R-t_n)/\hbar} C_\alpha |\Psi\rangle = \delta_{\beta\alpha}
R^{(C)}_{\alpha} e^{-iH(t_R-t_n)/\hbar} C_\alpha|\Psi\rangle\ .
\label{fourtwo}
\end{equation}
The relations obtained by summing (\ref{fourtwo}) over $\alpha$, or
alternatively over $\beta$, and using $\Sigma_\beta R_\beta = I$,
$\Sigma_\alpha C_\alpha = U(t,t_0)$ show that
\begin{equation}
R^{(C)}_\beta e^{-iH(t_R-t_0)/\hbar} |\Psi\rangle =
e^{-iH(t_R-t_n)/\hbar} C_\alpha
|\Psi\rangle\ .
\label{fourthree}
\end{equation}
The probabilities of the records of measurement outcomes are
therefore the same as the
probabilities of the histories
\begin{equation}
p^{(DH)}_{\alpha} \equiv \left\Vert C_\alpha |\Psi\rangle \right\Vert^2 =
\left\Vert R^{(C)}_\alpha e^{-iH(t_R-t_0)/\hbar} |\Psi\rangle \right\Vert^2
\ .
\label{fourfour}
\end{equation}

The situation in Bohmian mechanics is analogous with $C_\alpha$
replaced by $B_\alpha$, $R^{(C)}_\alpha$ by $R^{(B)}_\alpha$,
$p^{(DH)}_\alpha$ by $p^{(BM)}_\alpha$, {\it etc.} The analogs
of (\ref{fourtwo}) and (\ref{fourfour}) are
\begin{equation}
R^{(B)}_\beta e^{-iH(t_R-t_0)/\hbar} B_\alpha |\Psi\rangle=
\delta_{\alpha\beta}
R^{(B)}_\alpha e^{-iH(t_R-t_0)/\hbar} B_\alpha |\Psi\rangle\ .
\label{fourfive}
\end{equation}
and
\begin{equation}
p^{(B)}_\alpha \equiv \left\Vert B_\alpha |\Psi\rangle \right\Vert^2 =
\left\Vert R^{(B)}_\alpha e^{-iH(t_R-t_0)/\hbar} |\Psi\rangle \right\Vert^2\ .
\label{foursix}
\end{equation}

However, when the predictions of Bohmian mechanics and decoherent
histories differ, (\ref{fourtwo}) and (\ref{fourfour}) cannot be both
true with the same set of records $R^{(C)}_\alpha=R^{(B)}_\alpha
\equiv R_\alpha$ because that would imply the equality of
probabilities through (\ref{fourthree}) and (\ref{fourfive}).  
For instance, in the sequence of measurements described above, the
records of outcomes $\{R_\alpha\}$ can correlated either with the histories
$\{C_\alpha\}$ or the Bohmian trajectories $\{B_\alpha\}$ but not
both if $p^{(DH)}_\alpha \ne p^{(BM)}_\alpha$ 
The situations disussed in \cite{Engsum}   
where a detector localized in one region of space
registers a particle whose Bohmian trajectory is  
elsewhere are examples. The devices that measure 
position are not recording the position of the Bohmian trajectory.  
Conversely, in these situations different apparatus with different
records would be needed to measure the histories or the Bohmian
trajectories.

\section{The Past in Quantum Cosmology}
\label{sec: V}

Probabilities for the records of measurement outcomes are not the
only probabilities of interest in physics. For example, in
cosmology, and in other areas of inquiry, the probabilities of 
past history are central to our understanding of the present. 

Quantum mechanically, past history is a sequence of past events that is 
correlated with our present records with high probability. Why bother 
calculating these probabilities and using them to reconstruct the
past? It's over and done with. Reconstructing the past is useful because
it simplifies the prediction of the future. (See, {\it e.g.}
\cite{Har98b}). Take, for example, our understanding of the history
of the very early universe from which we predict the present large
scale distribution of the galaxies and the abundances of the
elements in parts of the universe as yet unseen. We do not measure
this early history. We infer with high probability from present 
observations.  In principle, those same
predictions could be made from a theory of the initial condition and the
corpus of records of present instrumental observations, but it is much
easier to first reconstruct the universe's past history from these, and 
from that predict the future.

As the BESSW example makes clear, Bohmian mechanics and decoherent
histories could differ significantly in their accounts of the past.
That can be true even when 
approximate classical determinism holds in decoherent histories. Bohmian 
trajectories are not necessarily classical although they are always
deterministic. 

The BESSW example is very special. It is a simple model with one
free particle and a very particular initial condition. It does not
generalize naturally to many interacting particles moving in 
three-dimensions. Yet more realistic three-dimensional calculations
with initial states that are in a superposition of wave packets
whose centers follow classical orbits show similar non-classical
behavior for Bohmian trajectories \cite{Erlerun}.

The question of whether the Bohmian trajectories describing the realistic 
past of our universe behave classically or not is an interesting 
question for future investigation. As a starting point, however,
it is worth noting that it 
is unlikely that the initial wave function of the universe
assigns a definite position to each galaxy. To do so it would have to
encode the complexity of the present large scale distribution of galaxies.
Rather, a simple, discoverable,  initial condition might be expected to be 
a {\it superposition} of all possible configurations of initial
positions.  Then
the complexity of the present distribution of galaxies
arose  from chance accidents
over the course of the universe's history rather than deterministically
from a complex initial condition. Contemporary theories of the initial
condition, such as Hawking's wave function of the universe
\cite{HHaw85}, 
have this superposition character. 

Suppose the Bohmian trajectories arising from a realistic wave function
of the universe were to exhibit significant non-classical behavior in the
epochs where the corresponding coarse-grained decoherent histories 
behaved classically with high probability.  
Bohmian mechanics and decoherent
histories formulations of quantum mechanics would then agree on the 
record of every measurement outcome, but disagree on the fundamental
description of the past. If so they might differ in their utility 
for cosmology.

\acknowledgments
Discussions with T. Brun, T. Erler, S. Goldstein, R. Griffiths, 
B. Hiley, and A.P.A. Kent are
gratefully acknowledged. This  work was supported in part by 
NSF Grant  PHY-0070895.


\begin{thebibliography}{99}

\bibitem{BH93} D.~Bohm and B.J.~Hiley, {\sl The Undivided Universe},
Routledge, London, (1993).

\bibitem{Gri84}  R.B.~Griffiths,  {\sl  J. Stat. Phys.}, {\bf 36}, 219,
1984.

\bibitem{Omn94} R.~Omn\`es, {\sl Interpretation of Quantum Mechanics},
(Princeton University Press, Princeton, 1994).

\bibitem{GH90a} M.~Gell-Mann and J.B.~Hartle,  in {\sl Complexity, Entropy,
and the Physics of Information, SFI Studies in the Sciences of
Complexity}, Vol.  VIII, ed. by W. Zurek,  Addison Wesley, Reading, MA
(1990).

\bibitem{Ken96} A.P.A. Kent, in  
{\sl Bohmian Mechanics and Quantum Theory: An
Appraisal},  ed. by J. Cushing, A. Fine and S. Goldstein,  (Kluwer
Academic Press, Dordrecht, 1996), pp. 343-352, quant-ph/9511032.  

\bibitem{Gri99} R.B.~Griffiths, {\sl Phys.Lett.},  A261, 227 (1999);  
 quant-ph/9902059. 

\bibitem{HM00} B.J.~Hiley and O.J.E.~Maroney, quant-ph/0009056.  

\bibitem{GH94} M.~Gell-Mann and J.B.~Hartle,  {\it Time Symmetry and 
Asymmetry in Quantum Mechanics and Quantum Cosmology} in
{\sl Physical Origins of Time Asymmetry},
ed. by  J.~Halliwell, J.~Perez-Mercader, and W.~Zurek, Cambridge
University Press, Cambridge (1994) pp.~311-345; gr-qc/9304023.

\bibitem{Har03a} J.B.~Hartle, {\it Theories of Everything and
Hawking's Wave Function of the Universe}, in {\sl The Future of 
Theoretical Physics and Cosmology}, ed. by G.W.~Gibbons,
E.P.S.~Shellard, and S.J.~Rankin, Cambridge University Press, 
Cambridge, 2003, pp.38-50, gr-qc/0209104.

\bibitem{Har93a} J.B.~Hartle, in {\sl Directions in General Relativity,
Volume 1: A Symposium and Collection of Essays in honor of Professor
Charles W. Misner's 60th Birthday}, ed. by B.-L.~Hu,
M.P.~Ryan, and C.V.~Vishveshwara, Cambridge University Press, Cambridge
(1993). gr-qc/9210006.

\bibitem{Bel80} J.S. Bell, {\sl Intl. J.~Quant.~Chem}, Quantum
Chemistry Symposium 14, John Wiley, New York (1980), pp.~155--159
[Reprinted in J.S. Bell, {\sl Speakable and Unspeakable in Quantum
Mechanics}, Cambridge University Press, Cambridge (1987), p.111-116].

\bibitem{Engsum} B.-G.~Englert, M.W.~Scully, G.~S\"ussmann, and 
H.~Walther, {\sl Z.~Naturforsch}, {\bf 47a}, 1175 (1992); {\it ibid.},
{\sl Z.~Naturforsch}, {\bf 48a}, 1263 (1993).


\bibitem{DFGZ93} D.~D\"urr, W.~Fusseder,
S.~Goldstein, and N.~Zanghi, {\sl Z.~Naturforsch}, {\bf 48a}, 1261 (1993).

\bibitem{DHS93} C. Dewdney, L. Hardy, and E.J.~Squires, {\sl Phys.
Lett.} {\bf A184}, 6 (1993).

\bibitem{AV96} Y. Aharonov and L. Vaidman, in 
{\sl Bohmian Mechanics and Quantum Theory: An
Appraisal},  ed. by J. Cushing, A. Fine and S. Goldstein,  (Kluwer
Academic Press, Dordrecht, 1996). 

\bibitem{HCM00} B.J.~Hiley, R.E.~Callaghan and O.J.E. Maroney,
quant-ph/0010020. 

\bibitem{Har91a}  J.B.~Hartle, in
{\sl Quantum Cosmology and Baby Universes:  Proceedings of the 1989 
Jerusalem Winter School for Theoretical Physics}, ed. by ~S.~Coleman, 
J.B.~Hartle, T.~Piran, and S.~Weinberg, World Scientific, Singapore
(1991), Section II.10,  pp. 65-157.


\bibitem{Har98b} J.B.~Hartle,  
{\sl Physica Scripta}, {\bf T76}, 67--77 (1998); gr-qc/9712001.

\bibitem{Erlerun} T. Erler (unpublished).

\bibitem {HHaw85} J. Halliwell and S.W. Hawking, 
%{\it Origin Structure in the Universe,}
   Phys. Rev. D {\bf 31}, 1777, 1985.

%\bibitem{Haw84} S.W.~Hawking, {\sl Nucl.~Phys.~B}, {\bf 239}, 257 (1984).

%\bibitem{Galmergers} See {\it e.g.} 
%W.~Wu and W.~Keel, 
%{\sl Astron. J.}, {\bf 116}, 1513 (1998);
%W. Percival and L.~Miller, 
% {\sl MNRAS},
%{\bf 309}, 823 (1999); V.P.~Reshetnikov, 
%{\sl Astron.~Ap.}, {\bf 353}, 92 (2000). 


\end{thebibliography}
\end{document}